\documentstyle[preprint,aps]{revtex}  

\begin{document}

\title{Pairs of charged heavy-fermions from an 
SU(3)$_{L}\otimes$U(1)$_{N}$ model at $e^+ e^-$  colliders}

\author{J.\ E.\ Cieza Montalvo $^{*}$}

\address{Instituto de F\'{\i}sica, Universidade do Estado do Rio de
Janeiro, \\ Rua S\~ao Francisco Xavier 524, 20559-900 Rio de Janeiro, RJ, Brazil.}

\author{M. D. Tonasse $^{*}$}

\address{Instituto Tecnol\'ogico de Aeron\'autica, Centro T\'ecnico  
Aeroespacial, \\ 
Pra\c ca Marechal do Ar Eduardo Gomes 50, 12228-900 S\~ao Jos\'e dos Campos, SP, Brazil}


\maketitle

\begin{abstract} 
We investigate the production, backgrounds and signatures of pairs of charged heavy-fermions using the SU(3)$_L\otimes$U(1)$_N$ electroweak model in $e^+ e^-$ colliders (NLC and CLIC). We also analyze the indirect evidence for a boson $Z^{'}$.



\end{abstract}


\section{INTRODUCTION}

Although the standard electroweak model is very well successful explaining experimental data up to order 100 GeV, there are experimental results on the muon anomalous magnetic moment \cite{Bea01} and (solar and atmospheric) neutrinos \cite{Aea99}, which suggest no standard interpretation. Some other known experimental facts, such as the proliferation of
the fermion generation and their complex pattern of masses and mixing angles, are not predicted in the framework of the standard model. There are no theoretical explanation for the existence of several generations and for the values of the masses. It was established at the CERN $e^{+} e^{-}$ collider LEP that the number of light neutrinos is three \cite{lep}.

Many models, such as composite models \cite{af,bu1}, grand unified theories \cite{la}, technicolor models \cite{di}, superstring-inspired models \cite{e6}, mirror fermions \cite{maa} predict the existence of new particles with masses around of the scale of $1$ TeV. All these models consider the possible existence of a new generation of fermions.

Heavy-leptons are usually classified in four types: sequential leptons, paraleptons,  ortholeptons, and long-lived penetrating particle \cite{Gea00}. In this work we will study the type of heavy leptons which does not belong to any one mentioned above. Consequently, the existing experimental bounds on heavy-lepton parameters do not apply to them. This kind of heavy-leptons are predicted by  an electroweak model based on the SU(3)$_{C}\otimes$SU(3)$_{L}\otimes$U(1)$_{N}$ (3-3-1 for short) semi simple symmetry group \cite{PT93}. In this model the new leptons require not new generations, as occur in the most of the heavy-lepton models \cite{FH99,cie}. It is a chiral electroweak model whose left-handed charged heavy-leptons, which we denote by $P_a$ $=$ $E$, $M$ and $T$, together in association with the ordinary charged leptons and its respective neutrinos, are accommodate in SU(3)$_L$ triplets. So, we will study the production mechanism for these heavy-exotic leptons,  together with the exotic quarks and exotic neutrinos, in $e^{-} e^{+}$ colliders such as the Next Linear Collider (NLC) ($\sqrt{s} = 500$ GeV) and CERN Linear Collider (CLIC) ($\sqrt{s} = 1000$ GeV). 

The outline of this paper is the following. In Sec. II we describe the relevant features of the model. The luminosities of $\gamma \gamma$, $\gamma Z$ and $ZZ$ for $e^{-} e^{+}$ colliders are given in Sec. III. In Sec. IV we study the production of a pair of exotic-lepton. We summarize the results in Sec. V.

\section{Basic facts about the 3-3-1 heavy-lepton model}   
\label{secII}    

The most interesting feature of this class of models is the occurrence of anomaly cancellations, which is implemented only when the three fermion families are considered together and not family by family as in the standard model. This implies that the number of families must be a multiple of the color number and, consequently, the 3-3-1 model suggests a route towards the response of the flavor question \cite{PP92}. The model has also a great phenomenological interest since the related new physics can be expected in a scale near of the Fermi one \cite{MT02,CQ99}.\par
Let us summarize the most relevant points of the model (for details see Ref. 
\cite{PT93}). The left-handed leptons and quarks transform under the SU(3)$_L$ gauge group as the triplets   
\begin{mathletters}  
\begin{equation}  
\psi_{aL} = \left(\begin{array}{c}  \nu_{\ell_a} \\  \ell_a^\prime \\  P^\prime_a 
\end{array}\right)_L \sim \left({\bf 3}, 0\right), \quad Q_{1L} = \left(\begin{array}{c} 
u^\prime_1 \\ d^\prime_1 \\  J_1 \end{array}\right)_L \sim \left({\bf 3}, \frac{2}{3}\right), 
\quad Q_{\alpha L} = \left(\begin{array}{c} J^\prime_\alpha \\  u^\prime_\alpha \\   
d^\prime_\alpha \end{array}\right)_L \sim \left({\bf 3}^*, -\frac{1}{3}\right),  
\label{cont} 
\end{equation}\noindent  
where $P^\prime_a$ $=$ $E^\prime$, $M^\prime$, $T^\prime$ are the new leptons, $\ell^\prime_a$ 
$=$ $e^\prime$, $\mu^\prime$, $\tau^\prime$ and $\alpha$ = 2, 3. The $J_1$ exotic quark carries 
$5/3$ units of elementary electric charge while $J_2$ and $J_3$ carry $-$4/3 each. In Eqs. (\ref{quark}) the numbers 0, 2/3 and $-$1/3 are the U(1)$_N$ charges. Each left-handed charged fermion has its right-handed counterpart transforming as a singlet in the presence of the SU(3)$_L$ group, {\it i. e.},    
\begin{eqnarray}  
\ell'_R \sim \left({\bf 1}, -1\right), & \qquad & P'_R \sim \left({\bf 1}, 1\right), \qquad U'_R 
\sim \left({\bf 1}, 2/3\right), \\  D'_R \sim \left({\bf 1}, -1/3\right), & \qquad & J_{1R} \sim 
\left({\bf 1}, 5/3\right), \qquad J_{2,3R}' \sim \left({\bf 1}, -4/3\right).   
\end{eqnarray}
\label{quark}
\end{mathletters}\noindent 
We are defining $U = u, c, t$ and $D = d, s, b$. In order to avoid anomalies, one of the quark families must transforms in a different way with respect to the two others. In Eqs. (\ref{quark}) all the primed fields are linear combinations of the mass eigenstates. The charge operator is defined by   
\begin{equation}  
\frac{Q}{e} = \frac{1}{2}\left(\lambda_3 - \sqrt{3}\lambda_8\right) + N,  
\label{op} 
\end{equation}  
where the $\lambda$'s are the usual Gell-Mann matrices. We notice, however, that since $Q_{\alpha L}$ in Eqs. (\ref{cont}) are in antitriplet representation of SU(3)$_L$, the 
antitriplet representation of the Gell-Mann matrices must be used also in Eq. (\ref{op}) in order to get the correct electric charge for the quarks of the second and third generations.\par    
The three Higgs scalar triplets
\begin{equation}
\eta = \left(\begin{array}{c} \eta^0 \\  \eta_1^- \\  \eta_2^+ \end{array}\right) \sim \left({\bf 3}, 0\right), \quad \rho = \left(\begin{array}{c} \rho^+ \\  \rho^0 \\  \rho^{++}
\end{array}\right) \sim \left({\bf 3}, 1\right), \quad \chi =
\left(\begin{array}{c} \chi^- \\
\chi^{--} \\ \chi^0 \end{array}\right) \sim \left({\bf 3}, -1\right),
\label{higgs}\end{equation}
generate the fermion and gauge boson masses in the model. The neutral scalar fields develop the vacuum expectation values (VEVs)  $\langle\eta^0\rangle = v_\eta$, $\langle\rho^0\rangle = v_\rho$ and  $\langle\chi^0\rangle = v_\chi$, with $v_\eta^2 + v_\rho^2 = v_W^2 = (246 \mbox{ GeV})^2$. Neutrinos can get their masses from the $\eta^0$ scalar. A detailed scheme for Majorana mass generation for the neutrinos in this model is given in Ref. \cite{OY99}.\par
The pattern of symmetry breaking is   
\[ \mbox{SU(3)}_L \otimes\mbox{U(1)}_N \stackrel{\langle\chi\rangle}{\longmapsto}\mbox{SU(2)}_L\otimes\mbox{U(1)}_Y 
\stackrel{\langle\eta, \rho\rangle}{\longmapsto}\mbox{U(1)}_{\rm em}\] 
and so, we can expect $v_\chi \gg v_\eta, v_\rho$. The $\eta$ and $\rho$ scalar triplets give masses to the ordinary fermions and gauge bosons, while the $\chi$ scalar triplet gives masses to the new fermions and new gauge bosons.\par
Due the transformation properties of the fermion and Higgs fields under SU(3)$_L$ [see Eqs. (\ref{quark}) and (\ref{higgs})] the Yukawa interactions in the model are    
\begin{mathletters} 
\begin{eqnarray}
{\cal L}_\ell^Y & = & -G_{ab}\bar\psi_{aL}\ell^\prime_{bR}\rho - G^\prime_{ab}\bar\psi_{aL}^\prime P^\prime_{bR}\chi + \mbox{H. c.}, 
\label{yl} \\ 
{\cal L}_q^Y & = & \sum_a\left[\bar Q_{1L}\left(G_{1a}U^\prime_{a R}\eta + \tilde G_{1a}D^\prime_{a R}\rho\right) + \sum_\alpha\bar  Q_{\alpha L}\left(F_{\alpha a}U^\prime_{aR}\rho^* +  \tilde F_{\alpha a}D^\prime_{aR}\eta^*\right)\right] + \cr &&  
\sum_{\alpha\beta}F^J_{\alpha\beta}\bar Q_{\alpha L}J^\prime_{\beta R}\chi^* + G^J\bar  Q_{1L}J_{1R}\chi +  \mbox{H. c.}  \nonumber  \\  
\label{yq} 
\end{eqnarray}\label{weak}\end{mathletters}\noindent
The $G$'s, $F$'s and $\tilde F$'s are Yukawa coupling constants with $a, b = 1, 2, 3$ and $\alpha, \beta = 2, 3$. The interaction eigenstates which appear in Eqs. (\ref{weak}) can be transformed in the corresponding physical eigenstates by appropriated rotations. However, since the cross section calculation imply summation on flavors (see Sec. \ref{secIV}) and  the rotation matrix must be unitary, the mixing parameters have not essential effect for our purpose here. So, thereafter we suppress the primes notation for the interactions eigenstates.\par    
The gauge bosons consist of an octet $W^i_\mu$ $\left(i = 1, \dots, 8\right)$ associated with SU(3)$_L$ and a singlet $B_\mu$ associated with U(1)$_N$. The covariant derivatives are
\begin{equation}
{\cal D}_\mu\varphi_a = \partial_\mu\varphi_a+ i\frac{g}{2}\left(\vec W_\mu.\vec\lambda\right)^b_a\varphi_b + ig^\prime N_\varphi\varphi_aB_\mu,
\end{equation}
where $\varphi = \eta, \rho, \chi$. The model predicts single charged $\left(V^\pm\right)$, double charged $\left(U^{\pm\pm}\right)$ vector bileptons and a new neutral gauge boson $\left(Z^\prime\right)$ in addition to the charged standard gauge bosons $W^\pm$ and the neutral standard $Z$. We take from Ref. \cite{MT02} the trilinear interactions of the $Z^\prime\left(k_1\right)$ with the $V\left(k_2\right)^\pm$ and $U^{\pm\pm}\left(k_3\right)$, in the usual notation that all the quadrimoments are incoming in the vertex,
\begin{equation}
{\cal V}_{\lambda\mu\nu} = -i\frac{g}{2}\sqrt{\frac{3}{1 + 3t_W^2}}\left[\left(k_1 - k_2\right)_\lambda g_{\mu\nu} + \left(k_2 - k_3\right)_\mu g_{\nu\lambda} + \left(k_3 - k_1\right)_\nu g_{\lambda\mu}\right]
\end{equation}

where

\begin{equation}  
t_W^2 = \frac{\sin^2{\theta_W}}{1 - 4\sin^2{\theta_W}}.  
\label{tw}
\end{equation} 
The relevant neutral vector current interactions are    
\begin{mathletters}
\begin{eqnarray}
{\cal L}_{Z} & = & -\frac{g}{2\cos\theta_W}\left[a_L\left(f\right)\overline{f}\gamma^\mu\left(1 -  \gamma_5\right)f + a_R\left(f\right)\overline{f}\gamma^\mu\left(1 - \gamma_5\right)f\right]Z_\mu, 
\label{lz}\\
{\cal L}_{Z^\prime} & = & -\frac{g}{2\cos\theta_W}\left[a^\prime_L\left(f\right)\overline{f}\gamma^\mu\left(1 - \gamma_5\right)f + a^\prime_R\left(f\right)\overline{f}\gamma^\mu\left(1 - \gamma_5\right)f\right]Z^\prime_\mu, 
\label{lzl}\\
{\cal L}_{AP} & = & -e\bar P_a\gamma^\mu P_aA_\mu, \\
{\cal L}_{ZP} & = & -g\sin{\theta_W}\tan{\theta_W}\bar P_a\gamma^\mu P_aZ_\mu,
\label{lagrb}\\
{\cal L}_{Z^\prime P} & = &  -\frac{g\tan{\theta_W}}{2\sqrt{3}t_W}\bar P_a\gamma^\mu\left[3t_W^2 - 1 + \left(3t_W^2 + 1\right)\gamma_5\right]P_aZ^\prime_\mu,
\label{lagrc}\\
{\cal L}_{Zq} & = & -\frac{g}{4\cos{\theta_W}}\sum_a\bar q_a\gamma^\mu\left(v^a + a^a\gamma^5\right)q_aZ_\mu,
\label{lagrd}\\
{\cal L}_{Z^\prime q} & = & -\frac{g}{4\cos{\theta_W}}\sum_a\bar q_a\gamma^\mu\left(v^{\prime a} 
+ a^{\prime a}\gamma^5\right)q_aZ^\prime_\mu,
\label{lagre}
\end{eqnarray}
\label{lagr}
\end{mathletters}\noindent 
where $\theta_W$ is the Weinberg mixing angle, $f$ is any fermion and $q_a$ is any quark  \cite{PT93,PP92}. The coefficients in Eqs. (\ref{lz}), (\ref{lzl}), (\ref{lagrd}) and (\ref{lagre}) are   
\begin{mathletters}  
\begin{eqnarray}  
a_L\left(\nu_a^\prime\right) = \frac{1}{2}, \quad a_R\left(\nu_a\right) = 0, \quad a^\prime_L\left(\nu_a^\prime\right) & = & \frac{1}{2}\sqrt{\frac{1 - 4\sin^2\theta_W}{3}}, \quad a^\prime_R\left(\nu_a^\prime\right) = 0, \\
a_L\left(e_a^\prime\right) = -\frac{1}{2} + \sin\theta_W, \quad a_R\left(e_a^\prime\right) = \sin\theta_W, & \quad & a^\prime_L\left(e^\prime_a\right) = a^\prime_L\left(\nu^\prime_a\right), \quad a^\prime_R\left(e^\prime_a\right) = -\frac{\sin\theta_W}{2a^\prime_L\left(\nu^\prime_a\right)} \\
a_L\left(E^\prime_a\right) = a_R\left(E^\prime_a\right) = -\sin\theta_W, \quad a^\prime_L\left(E^\prime_a\right) & = & -\sqrt{\frac{1 - 4\sin\theta_W}{3}}, \quad a^\prime_R\left(E^\prime_a\right) = -a^\prime_R\left(e^\prime_a\right) \\
v^U = \frac{3 + 4t_W^2}{f\left(t_W\right)}, \quad v^D = -\frac{3 + 8t_W^2}{f\left(t_W\right)}, & \quad & -a^U = a^D = 1, \quad  v^{\prime u} = -\frac{1 + 
8t_W^2}{f\left(t_W\right)}, \\  
v^{\prime c} = v^{\prime t} = \frac{1 - 
2t_W^2}{f\left(t_W\right)}, \quad v^{\prime d} = -\frac{1 + 2t_W^2}{f\left(t_W\right)}, & \quad &
v^{\prime s} =  v^{\prime b} = \frac{f\left(t_W\right)}{\sqrt(3)}, \quad a^{\prime u} =  
\frac{1}{f\left(t_W\right)}, \\ 
a^{\prime c} = a^{\prime t} = -\frac{1 + 6t_W^2}{f\left(t_W\right)}, \quad a^{\prime d} = 
-a^{\prime c}, & \quad & a^{\prime s} = a^{\prime b} = -a^{\prime u} \quad v^{\prime J_1} = 
\frac{2\left(1 - 7t_W^2\right)}{f\left(t_W\right)}, \\ 
v^{\prime J_2} = v^{\prime J_3} = 
-\frac{2\left(1 - 5t_W^2\right)}{f\left(t_W\right)}, & \quad &
a^{\prime J_2} = a^{\prime J_2} = a^{\prime J_2} = a^{\prime J_1} = -\frac{2\left(1 + 
3t_W^2\right)}{f\left(t_W\right)},
\end{eqnarray} 
\label{coef}
\end{mathletters} 
with $f^2\left(t_W\right) = 3\left(1 + 4t_W^2\right)$. As we comment in Sec. I and by inspection of Eqs. (\ref{quark}), (\ref{lagrb}) and (\ref{lagrc}), we conclude that the heavy-leptons $P_a$ belong to another  class of exotic particles differently of the heavy-lepton classes usually considered in the literature. Thus, the  present experimental limits 
do not apply directly to them \cite{Gea00} (see also Ref. \cite{PT93}). Therefore, the 3-3-1 heavy-leptons phenomenology deserves more detailed studies.\par    

\section{LUMINOSITIES}    

We analyze the case of elastic $e^{-} e^{+}$ scattering. The $\gamma \gamma$ differential luminosity is given by 

\begin{eqnarray}
\left (\frac{d \rm L^{el}}{d\tau} \right )_{\gamma \gamma/\ell \ell}  =  \int_{\tau}^{1} \frac{dx_{1}}{x_{1}} f_{\gamma/\ell} (x_{1})    
f_{\gamma/\ell} (x_{2} = \tau/x_{1})   \; ,
\end{eqnarray}  
where  $\tau = x_{1} x_{2}$ and $f_{\gamma/\ell} (x)$ is the effective photon approximation for the  photon into the lepton, which is defined by

\begin{eqnarray}
f_{\gamma/\ell} (x) = \frac{\alpha}{2\pi} \frac{1+ (1- x)^{2}}{x} \ln \frac{s}{4 m_{e}^{2}} \; ,
\nonumber 
\end{eqnarray} 
where x is the longitudinal momentum fractions of the lepton carried off by the photon, $s$ is the center-of-mass energy of the $e^{-} e^{+}$ pair and $m_{e}$ is the electron mass.

The $ZZ$ differential luminosity for elastic $e^{-} e^{+}$ scattering is given by 

\begin{eqnarray}
\left (\frac{d \rm L^{el}}{d\tau} \right )_{Z Z/\ell \ell}  = \int_{\tau}^{1} \frac{dx_{1}}{x_{1}} f_{Z/\ell} (x_{1})    
f_{Z/\ell} (x_{2} = \tau/x_{1})   \; ,
\nonumber
\end{eqnarray} 
where $f_{Z/\ell}(x)$ is the distribution function for finding a boson Z of transverse and longitudinal helicities in a fermion with energy $\sqrt{s}$ in the limit $\sqrt{s} \ge 2M_{Z}$ and which have the following forms

\begin{eqnarray}
f^{\pm T}_{Z/\ell} (x) = \frac{\alpha}{4 \pi x \sin^{2}{\theta_{W}} \cos^{2}{\theta_{W}}} \left [ \left ( g_{V}^{\ell} \mp g_{A}^{\ell} \right )^{2} +  \left ( g_{V}^{\ell} \pm g_{A}^{\ell} \right )^{2} (1- x)^{2} \right ] \ln \frac{s}{M_{Z}^{2}} \; ,
\nonumber 
\end{eqnarray}  

\begin{eqnarray}
f^{L}_{Z/\ell} (x) = \frac{\alpha}{\pi \sin^{2}{\theta_{W}} \cos^{2}{\theta_{W}}} \left [ \left ( g_{V}^{\ell} \right )^{2} +  \left( g_{A}^{\ell} \right )^{2} \right ]  \frac{1- x}{x}  \; ,
\nonumber 
\end{eqnarray}  
where the $g_{V}^{\ell}$ and $g_{A}^{\ell}$ are the vector and axial-vector coupling.

And for the $Z \gamma$ differential luminosity for elastic $e^{-} e^{+}$ scattering we have 

\[
\left (\frac{d \rm L^{el}}{d\tau} \right )_{Z \gamma /\ell \ell}  = \int_{\tau}^{1} \frac{dx_{1}}{x_{1}} f_{Z/\ell} (x_{1}) f_{\gamma /\ell} (x_{2} = \tau /x_{1}) 
\]


\section{CROSS SECTION PRODUCTION}

\label{secIV}

\subsection{$e^{-} e^{+} \rightarrow P^{-} P^{+}$}

Pair production of exotic particles is, to a very good
approximation, a model independent process, since it proceeds through a well known electroweak interaction. This production mechanism can be  studied through the analysis of the reactions $e^{-} e^{+} \rightarrow P^{-}  P^{+}$, provided that there is enough available energy ($\sqrt{s} \geq 2M_{P}$). We will analyze the following processes for pair production of exotic heavy leptons:   $e^{-} e^{+} \rightarrow  P^{-} P^{+}$, $e^{-} e^{+} \rightarrow \gamma \gamma \rightarrow P^{-}  P^{+}$, $e^{-} e^{+} \rightarrow Z \gamma \rightarrow P^{-}  P^{+}$ and $e^{-} e^{+} \rightarrow  Z Z \rightarrow P^{-}  P^{+}$, the first process take place through the exchange of a photon, a  boson $Z^{0}$ and ${Z^{0}}^{'}$ in the $s$ channel, while the others processes take place through the exchange of heavy lepton in the $t$ and $u$ channel.

Using the interactions Lagrangians (8a),(8b) and (8c), it is easy to evaluate the cross section for the process $e^{+} e^{-} \rightarrow P^{+} P^{-}$, involving a neutral current, from which we obtain:

\begin{eqnarray} 
\left (\frac{d \sigma}{d\cos \theta} \right )_{P^+P^-} =  &&\frac{\beta 
\alpha^{2} \pi} {s^{3}} \Biggl \{ \left[ 2 s M_P^{2}
+ \left(M_P^{2} - t\right)^{2} + \left(M_P^{2} - u\right)^{2}  \right]  
\nonumber \\
&&+ \frac{1}{2 \sin^{2} \theta_{W} \cos^{2} \theta_{W} \left(s -
M_{Z,Z'}^{2} + i M_{Z,Z'} \Gamma_{Z,Z'}\right)} 
\left[ 2s M_P^{2} {g^\prime}_{V}^{PP} g_{V}^{\ell}\right. \nonumber \\ 
&& \left. + {g^\prime}_{V}^{PP} g_{V}^{\ell} \left[\left(M_P^{2} - t\right)^{2} + 
\left(M_P^{2} - u\right)^{2}\right] + {g^\prime}_{A}^{PP} g_{A}^{\ell} \left( \left(M_P^{2} - u\right)^{2} \right.\right. \nonumber \\ && \left.\left. -\left(M_P^{2} - t\right)^{2}\right) \right]  \Biggr \} + \frac{\beta \pi \alpha^{2}}{16 \cos^{4} \theta_{W} \sin^{4} \theta_{W}} \frac{1}{s \left(s - M_{Z,Z'}^{2} + i M_{Z,Z'} \Gamma_{Z,Z'}\right)^{2}} \nonumber \\
&& \times\Biggl \{\left[\left({g^\prime}_V^{PP}\right)^2 + 
\left({g^\prime}_{A}^{PP}\right)^{2}\right]\left[\left(g_V^\ell \right)^2 +  
\left(g_{A}^{\ell}\right)^{2}\right] \left[\left(M_P^{2} - u\right)^{2}+\left(M_P^{2} - t\right)^{2} \right]  \nonumber \\
&& + 2 s M_{P}^{2} \left[\left({g^\prime}_{V}^{PP}\right)^{2} - 
\left({g^\prime}_{A}^{PP}\right)^{2}\right]\left[\left(g_{V}^{\ell}\right)^{2} + 
\left(g_{A}^\ell \right)^{2}\right]   \nonumber   \\
&& + 4{g^\prime}_{V}^{PP}{g^\prime}_{A}^{PP} g_{V}^{\ell} g_{A}^{\ell} 
\left[\left(M_P^{2} - u\right)^{2} - \left(M_P^{2} - t\right)^{2}\right] \Biggr 
\}  \nonumber   \\
&&+ \frac{\beta \pi \alpha^{2}}{8 \sin^{4} \theta_{W} \cos^{4} 
\theta_{W} s\left(s- M_{Z}^{2} + i M_{Z} \Gamma_{Z}\right) \left(s- M_{Z'}^{2}+ i M_{Z'} \Gamma_{Z'}\right)}   \nonumber    \\
&&\Biggl \{2sM_{P}^{2} \left(g_{V}^{\ell} + g_{A}^{\ell}\right) 
\left(g_{V}^{PP}{g^\prime}_{V}^{PP} - g_{A}^{PP}{g^\prime}_{A}^{PP}\right) 
+ \left(M_{P}^{2}- t\right)^{2}  \left[  \left( \left(g_{V}^{\ell}\right)^{2} + 
\left(g_{A}^{\ell}\right)^{2}\right)\right.  \nonumber   \\  
&& \left. g_{V}^{PP}{g^\prime}_{V}^{PP} + g_{A}^{PP}{g^\prime}_{A}^{PP} - 2g_{V}^{\ell} g_{A}^{\ell} g_{V}^{PP}{g^\prime}_{A}^{PP} - 
2g_{V}^{\ell}g_{A}^{\ell}g_{A}^{PP}{g^\prime}_{V}^{PP}\right]  \nonumber \\
&& + \left(M_{P}^{2}- u\right)^{2}  \left[  \left( \left(g_{V}^{\ell}\right)^{2} + 
\left(g_{A}^{\ell}\right)^{2}\right) 
\left(g_{V}^{PP}{g^\prime}_{V}^{PP} + g_{A}^{PP}{g^\prime}_{A}^{PP}\right)\right.  \nonumber \\
&& \left. + 2g_{V}^{\ell} g_{A}^{\ell} g_{V}^{PP}{g^\prime}_{A}^{PP} + 2 
g_{V}^{\ell}g_{A}^{\ell} g_{A}^{PP}{g^\prime}_{V}^{PP}\right] \Biggr\},
\end{eqnarray}\\
where 

\[
g_{V,A}^{PP} = \frac{a_L \pm a_R}{2}, \qquad {g^\prime}_{V,A}^{PP} = 
\frac{a^\prime_{L} \pm a^\prime_{R}}{2}.
\]

The primes $\left(^\prime\right)$ is for the case when we take a boson $Z'$, $\Gamma_{Z,Z'}$ are the total width of the boson Z and $Z'$ \cite{MT02}, $\beta = \sqrt{1 - 4 M_P^{2}/s}$ is the velocity of the heavy-lepton in the c. 
m. of the process, $\alpha$ is the fine structure constant, which we take equal to $\alpha =1/128$, $g^{\ell}_{V, A}$ are the standard coupling constants, $M_{Z}$ is the mass of the $Z$ boson, $\sqrt{s}$ is the center of mass energy of the $e^{-} e^{+}$  system, $t = M_{P}^{2} - (1 - \beta \cos \theta)s/2$ and $u = M_{P}^{2} - (1 + \beta \cos \theta)s/2$, where $\theta$ is the angle between the heavy-lepton and the incident electron, in the c. m. frame. For $Z^\prime$ boson we take $M_{Z^\prime} = \left(0.6 - 3\right)$ TeV, since $M_{Z^\prime}$ is 
proportional to VEV $v_\chi$ \cite{PP92,FR92}. For the standard model parameters we assume PDG values, {\it i. e.}, $M_Z = 91.02$ GeV, $\sin^2{\theta_W} = 0.2315$ and $M_W = 80.33$ GeV \cite{Gea00}.\par

Another way to produce a pair of heavy exotic leptons is through the  elastic reactions of the type $e^{-} e^{+} \rightarrow \gamma \gamma   \rightarrow P^{-} P^{+}$, $e^{-} e^{+} \rightarrow Z \gamma \rightarrow P^{-} P^{+}$ and $e^{-} e^{+} \rightarrow Z Z   \rightarrow P^{-} P^{+}$ 

The three processes take place through the exchange of the exotic lepton in the $t$ and $u$ channels. So the cross section for the production of a pair of $P^{-} P^{+}$ in the $e^{-}  e^{+}$ collisions, we obtained by convoluting the cross section for the subprocess $\gamma \gamma \rightarrow P^{-} P^{+}$, $Z \gamma \rightarrow P^{-} P^{+}$ and $Z Z \rightarrow P^{-} P^{+}$, with the two photon, Z$\gamma$ and ZZ luminosities in this collisions, that is

\[
\sigma = \int_{\tau_{min}}^{1} \frac{d \it{L}}{d \tau} d\tau \ \hat{\sigma} (\hat{s} = x_{1} x_{2} s) = \int_{\tau_{min}}^{1}  \int_{\ln{\sqrt(\tau)}}^{-\ln{\sqrt(\tau)}} \frac{dx_{1}}{x_{1}} f_{V/\ell} (x_{1}) f_{V/\ell} (x_{2}) \int \frac{d \hat{\sigma}}{dcos} dcos 
\]
where $V = \gamma, Z$. The subprocess cross section for two photon $P^{-} P^{+}$ production via elastic collisions of electron-positron is 

\begin{eqnarray} 
\left (\frac{d \sigma}{d\cos \theta} \right )_{\gamma \gamma} = &&\frac{\beta \alpha^{2} \pi}{s} \bigl [\frac{1}{(t - M_{P}^{2})^{2}} (-M_{P}^{4}- 3 M_{P}^{2} t - M_{P}^{2} u + tu)  \nonumber \\
&&+ \frac{1}{(u - M_{P}^{2})^{2}} (-M_{P}^{4}- M_{P}^{2} t - 3 M_{P}^{2} u + tu) \nonumber  \\ 
&&+ \frac{2}{(t - M_{P}^{2}) (u - M_{P}^{2})} (-2 M_{P}^{4} - M_{P}^{2} t - M_{P}^{2} u ) \bigr ]  \; ,
\end{eqnarray}
where $M_{P}$ is the mass of the exotic lepton, $\hat{t} = M_{P}^{2}- \frac{\hat{s}}{2} (1- \beta cos \theta)$ and $\hat{u} = M_{P}^{2}- \frac{\hat{s}}{2} (1+ \beta cos \theta)$ refer to the exchanged momenta squared, corresponding to the direct and crossed diagrams for the two photon, with $\beta$ being the $P$ velocity in the subprocess c.m. and $\theta$ its angle with respect to the incident electron in this frame.

The contribution of the subprocess cross section for $Z \gamma$ and $Z Z$ luminosities to the total cross section is so small that we not include here the explicit calculation. Even so, we present the results in Fig. $3$ for the CLIC.


\subsection{$e^{-} e^{+} \rightarrow Q \bar{Q}$}

The production of exotic quarks was already studied by both of the autors \cite{cie,tonpe}, so that in this subsection we will study it through the analysis of the reaction $e^{-} e^{+} \rightarrow Q \bar{Q}$, provided that there is enough available energy ($\sqrt{s} \ge 2 M_{Q}$). Such process takes place through the exchange of a photon, Z and a $Z^{'}$ in the s channel.

Using the interaction Lagrangians, given by Sec. II, we can evaluate the cross section  involving a neutral current and obtain

\begin{eqnarray} 
\left (\frac{d \sigma}{d\cos \theta} \right )_{Q \bar{Q}} =  &&\frac{N_{c} \beta_{Q} 
\alpha^{2} \pi} {s^{3}} \Biggl \{ \left[ c_{q}^{2} (2 s M_Q^{2}
+ \left(M_Q^{2} - t\right)^{2} + \left(M_Q^{2} - u\right)^{2} )  \right]  
\nonumber \\
&&+ \frac{c_{q}}{2 \sin^{2} \theta_{W} \cos^{2} \theta_{W} \left(s -
M_{Z,Z'}^{2} + i M_{Z,Z'} \Gamma_{Z,Z'}\right)} 
\left[ 2s M_Q^{2} {g^\prime}_{V}^{Q \bar{Q}} g_{V}^{\ell}\right. \nonumber \\ 
&& \left. + {g^\prime}_{V}^{Q \bar{Q}} g_{V}^{\ell} \left[\left(M_Q^{2} - t\right)^{2} + 
\left(M_Q^{2} - u\right)^{2}\right] + {g^\prime}_{A}^{Q \bar{Q}} g_{A}^{\ell} \left( \left(M_Q^{2} - u\right)^{2} \right.\right. \nonumber \\ && \left.\left. -\left(M_Q^{2} - t\right)^{2}\right) \right]  \Biggr \} + \frac{\beta \pi \alpha^{2}}{16 \cos^{4} \theta_{W} \sin^{4} \theta_{W}} \frac{1}{s \left(s - M_{Z,Z'}^{2} + i M_{Z,Z'} \Gamma_{Z,Z'}\right)^{2}} \nonumber \\
&& \times\Biggl \{\left[\left({g^\prime}_V^{Q \bar{Q}}\right)^2 + 
\left({g^\prime}_{A}^{Q \bar{Q}}\right)^{2}\right]\left[\left(g_V^\ell \right)^2 +  
\left(g_{A}^{\ell}\right)^{2}\right] \left[\left(M_Q^{2} - u\right)^{2}+\left(M_Q^{2} - t\right)^{2} \right]  \nonumber \\
&& + 2 s M_{Q}^{2} \left[\left({g^\prime}_{V}^{Q \bar{Q}}\right)^{2} - 
\left({g^\prime}_{A}^{Q \bar{Q}}\right)^{2}\right]\left[\left(g_{V}^{\ell}\right)^{2} + 
\left(g_{A}^\ell \right)^{2}\right]   \nonumber   \\
&& + 4{g^\prime}_{V}^{Q \bar{Q}}{g^\prime}_{A}^{Q \bar{Q}} g_{V}^{\ell} g_{A}^{\ell} 
\left[\left(M_Q^{2} - u\right)^{2} - \left(M_Q^{2} - t\right)^{2}\right] \Biggr 
\}  \nonumber   \\
&&+ \frac{\beta \pi \alpha^{2}}{8 \sin^{4} \theta_{W} \cos^{4} 
\theta_{W} s\left(s- M_{Z}^{2} + i M_{Z} \Gamma_{Z}\right) \left(s- M_{Z'}^{2}+ i M_{Z'} \Gamma_{Z'}\right)}   \nonumber    \\
&&\Biggl \{2sM_{Q}^{2} \left(g_{V}^{\ell} + g_{A}^{\ell}\right) 
\left(g_{V}^{Q \bar{Q}}{g^\prime}_{V}^{Q \bar{Q}} - g_{A}^{Q \bar{Q}}{g^\prime}_{A}^{Q \bar{Q}}\right) 
+ \left(M_{Q}^{2}- t\right)^{2}  \left[  \left( \left(g_{V}^{\ell}\right)^{2} + 
\left(g_{A}^{\ell}\right)^{2}\right)\right.  \nonumber   \\  
&& \left. g_{V}^{Q \bar{Q}}{g^\prime}_{V}^{Q \bar{Q}} + g_{A}^{Q \bar{Q}}{g^\prime}_{A}^{Q \bar{Q}} - 2g_{V}^{\ell} g_{A}^{\ell} g_{V}^{Q \bar{Q}}{g^\prime}_{A}^{Q \bar{Q}} - 
2g_{V}^{\ell}g_{A}^{\ell}g_{A}^{Q \bar{Q}}{g^\prime}_{V}^{Q \bar{Q}}\right]  \nonumber \\
&& + \left(M_{Q}^{2}- u\right)^{2}  \left[  \left( \left(g_{V}^{\ell}\right)^{2} + 
\left(g_{A}^{\ell}\right)^{2}\right) 
\left(g_{V}^{Q \bar{Q}}{g^\prime}_{V}^{Q \bar{Q}} + g_{A}^{Q \bar{Q}}{g^\prime}_{A}^{Q \bar{Q}}\right)\right.  \nonumber \\
&& \left. + 2g_{V}^{\ell} g_{A}^{\ell} g_{V}^{Q \bar{Q}}{g^\prime}_{A}^{Q \bar{Q}} + 2 
g_{V}^{\ell}g_{A}^{\ell} g_{A}^{Q \bar{Q}}{g^\prime}_{V}^{Q \bar{Q}}\right] \Biggr\},
\end{eqnarray}\\
where $\beta_{Q} = \sqrt{1- 4 M_{Q}^{2}/s}$ is the velocity of the 
exotic-quark in the c.m. of the process, $c_{q}$ is the charge of the quark, $Q$ is the exotic quark and $\bar{Q}$ the exotic antiquark,  $\sqrt{s}$ is the center of mass energy of the $e^{-} e^{+}$ system, $t = M_{Q}^{2} - \frac{s}{2} (1 - \beta \cos \theta)$ and {} $u = M_{Q}^{2} - \frac{s}{2} (1 + \beta \cos \theta)$,  where $\theta$ is the angle between the exotic quark and the incident eletron, in the c.m. frame and the couplings $g_{V}^{QQ}$ and  $g_{A}^{QQ}$ are given in section II.

In other to analyze the indirect evidence for a boson $Z^{'}$, we calculate the production of the quarks as in the standard model as in the $3-3-1$ model, as a result we found that at high energies in the $3-3-1$ model there will be many more dijets than expected in the scope of the standard model. In Fig. $4$, we exhibit the cross section $\sigma(e^{-} e^{+} \rightarrow q \bar{q} (Q \bar{Q}))$ as a function of center of mass energy for diferent values of the boson mass $M_{Z^{'}}$.


\subsection{$e^{-} e^{+} \rightarrow N_{1} N_{2}$}

In this section we will study the massive neutrinos, due that there are indications for the existence of it mass. This ones are the deficit of solar electron neutrinos which flux is below that predicted by the standard solar model \cite{bahc}, neutrino oscillations, where the electron neutrinos partially convert to muon neutrinos within the interior of the sun \cite{mikh} and the need for explications the hot dark matter \cite{cald}. 

The production of massive neutrinos we study through the analysis of the reaction $e^{-} e^{+} \rightarrow N_{1} N_{2}$, this process takes place through the exchange of the bosons $Z, Z^{'}$ in the s channel. Using the interactions Lagrangians of the Eq. $..$, we evaluate the cross section obtaining

\begin{eqnarray} 
\left (\frac{d \sigma}{d\cos \theta} \right )_{N_{1} N_{2}} = &&\frac{\beta_{N} \alpha^{2} \pi}{32 s \sin^{4} \theta_{W} \cos^{4} \theta_{W}} \bigl [\frac{1}{\left(s -M_{Z,Z'}^{2} + i M_{Z,Z'} \Gamma_{Z,Z'}\right)} \bigl (2 M_{N}^{4} \left ({g_{V}^{\ell}}^{2} + {g_{A}^{\ell}}^{2} \right )  \nonumber  \\
&&- 2 M_{N}^{2} t \left (g_{V}^{\ell}- g_{A}^{\ell} \right)^{2}  
- 2 M_{N}^{2} u \left (g_{V}^{\ell} + g_{A}^{\ell} \right )^{2} + t^{2} \left (g_{V}^{\ell}- g_{A}^{\ell} \right)^{2} + u^{2} \left (g_{V}^{\ell} + g_{A}^{\ell} \right )^{2}  \bigr )  \nonumber \\
&& +\frac{2}{\left(s- M_{Z}^{2} + i M_{Z} \Gamma_{Z}\right) \left(s- M_{Z'}^{2}+ i M_{Z'} \Gamma_{Z'}\right)} \bigl (2 M_{N}^{4} (g_{V}^{\ell} g_{V}^{\ell'} + g_{A}^{\ell} g_{A}^{\ell'})  \nonumber  \\
&&- 2 M_{N}^{2} g_{V}^{\ell} (g_{V}^{\ell'}- g_{A}^{\ell'}) (t - u)+ 2 M_{N}^{2} g_{A}^{\ell} t  (g_{V}^{\ell'}- g_{A}^{\ell'}) - 2 M_{N}^{2} g_{A}^{\ell} u (g_{V}^{\ell'}+ g_{A}^{\ell'})  \nonumber  \\
&&+ t^{2} (g_{V}^{\ell'}- g_{A}^{\ell'}) (g_{V}^{\ell}- g_{A}^{\ell}) + u^{2} (g_{V}^{\ell'}+ g_{A}^{\ell'}) (g_{V}^{\ell}+ g_{A}^{\ell}) \bigr ]  \; ,
\end{eqnarray}
where $\beta_{N} = \sqrt{1- 4 M_{N}^{2}/s}$ is the velocity of
exotic-neutrino in the c.m. of the process, $M_{Z,Z'}$ is the mass of
the boson $Z(Z')$.


\section{RESULTS AND CONCLUSIONS}

In the following Figures we present the cross section for the process
$e^{+} e^{-} \rightarrow P^{+} P^{-}, (\bar{Q} Q), (N_{1} N_{2})$ for the NLC and CLIC. In all calculations we take $\sin^{2} {\theta_W} = 0.2315$, $M_Z = 91.188$ GeV, and the mass of the heavy exotic lepton equal to $200$ GeV

In Fig. $1$, we exhibit the cross section $\sigma (e^{-} e^{+} \rightarrow P^{-} P^{+})$ as a function of $M_{P}$. Taking into account that the expected integrated luminosity for the NLC, will be of order of $6 \times 10^{4} pb^{-1}/yr$, there will a total of: $\simeq 2,5 \times 10^{4}$ heavy exotic leptons pairs produced per year, considering $M_{Z}^{'} = 1200$ GeV, while for the $M_{Z'} = 2000$ GeV the production will be of order of $2,2 \times 10^{4}$.

In Fig. $2$, taking into account that the integrated luminosity for the CLIC will be of order of $2 \times 10^{5} pb^{-1}/yr$, then the statistics that we can expect for this collider is a little larger. So for the process $e^{-} e^{+} \rightarrow P^{-} P^{+}$, considering the mass of the boson $Z^{'}$ equal to $1200$ GeV, we will have a total of $\simeq 2 \times 10^{5}$ lepton pairs produced per year, while for $M_{Z'}= 2000$ GeV will be $\simeq 3 \times 10^{4}$, respectively. For both figures was taken $M_{J_{1}} = 300$ GeV, $M_{J_{2}} = 400$ GeV, $M_{J_{3}} = 600$ GeV and $M_{V} = 800$ GeV.

In Fig. $3$, we show the pair production of exotic heavy leptons through the elastic reactions, so the statistics that we can expect for the CLIC collider, for photon-photon $P^{-} P^{+}$ production, will be of order of $\simeq 2 \times 10^{4}$ lepton pairs produced per year, while for the $Z \gamma$ will be of $\simeq 200$ events per year and for the $ZZ$ will be very small. It is to note here that it was taken only the transverse helicity of the boson Z, since the longitudinal ones gives a small contribution.

In Fig. $4$, we compare the standard cross section $\sigma (e^{-} e^{+} \rightarrow q \bar{q})$ \ with the productions cross section $\sigma (e^{-} e^{+} \rightarrow q \bar{q} + Q \bar{Q})$, when the $3-3-1$ model is applied. We see from these results that using the $3-3-1$ model we will have more dijets than using the standard model at high energies. This figure was obtained imposing the cut $|\cos \theta| < 0.95$ and assuming three bosons $Z'$ with masses equal to $800$, $1200$ and $2000$ GeV respectively. This figure still exhibit the resonance peaks associated with the boson $Z'$. Here was taken for this figure $M_{J_{1}} = 200$ GeV, $M_{J_{2}} = 220$ GeV, $M_{J_{3}} = 245$, whose masses would be accessible to the NLC.

In Figs. $5$ and $6$ we show the cross sections for the production of
exotic quarks $e^{+} e^{-} \rightarrow \bar{Q} Q$, in the colliders NLC and CLIC. We see from these
results that we can expect for the first collider a total of $\simeq
3,6 \times 10^{5}$ heavy quark pairs produced per year, considering the mass of the boson $Z^{'}$ equal to $1.2$ and $2$ TeV. We see that the cross section for both masses of the boson $Z^{'}$ is not different one from another. For the second collider, that is for the CLIC we expect a total of $\simeq 2,4 \times 10^{6}$ exotic quarks for the mass of the boson $Z^{'}$ equal to $1,2$ TeV, while for $M_{Z}^{'} = 2$ TeV we obtain $4 \times 10^{5}$ events per year. Here the cross sections are different one from another, which is not the case for the NLC, this is due to the propagator, that for the CLIC is larger than for the NLC.

In Figs. $7$ and $8$ we show the cross section for the production of
exotic neutrinos, $e^{+} e^{-} \rightarrow N_{1} N_{2}$, in the colliders NLC and CLIC. We see from these results that we can expect, in the NLC, a total of  around $1,5 \times 10^{3}$ heavy neutrinos pairs produced per year for the mass of the boson $Z^{'}$ equal to $1,2$ TeV, while for the mass equal to $M_{Z}^{'}= 2$ TeV, the total of events is $1,3 \times 10^{3}$. We see that the cross sections are nearly equal. We also have that the CLIC can produce a total of $2 \times 10^{4}$ pairs of exotic neutrinos for the mass of the boson $Z^{'}$ equal to $1200$ GeV, while for $M_{Z}^{'} = 2$ TeV the number of events will be $5,8 \times 10^{3}$. The discrepancy between these  cross sections in both colliders, has the same reason as above. Here for both figures was taken $M_{J_{1}} = 300$ GeV, $M_{J_{2}} = 400$ GeV, $M_{J_{3}} = 600$ GeV and $M_{V} = 800$ GeV.

The main background for the signal, $e^{-} e^{+} \rightarrow P^{-}  P^{+} \rightarrow \bar{\nu} \bar{u} J_{1} \ (\nu u \bar{J_{1}} )$, could comes from the process $e^{-} e^{+} \rightarrow W^{-} W^{+} (Z \ Z) \rightarrow q \bar{q'} q \bar{q'} (q \bar{q} q \bar{q} )$, but this background can be eliminated by measuring the undetected neutrino contribution $p_{\nu T}$, since all hadrons with appreciable $p_{T}$ are detected. Then the overall $p_{T}$ imbalance for detected particles gives this undetected $p_{\nu T}$. Another background such as $e^{-} e^{+} \rightarrow W^{-} W^{+} (Z \ Z) \rightarrow e \nu q \bar{q'} (e^{+} e^{-} q \bar{q} )$, also can be eliminated, then the number of jets are different.\par

There is no standard backgrounds for the signal, $e^{-} e^{+} \rightarrow Q \bar{Q} \rightarrow q \ell^{-} \ell^{-} (\bar{q} \ell^{+} \ell^{+})$, due that in the standard model there is not leptonic number violation \cite{tonpe}, and the backgrounds for heavy neutrinos may be seen in \cite{sim}.

In summary, we showed in this work that in the context of the 3-3-1 model the signatures for heavy-fermions can be significant as in NLC as in the CLIC collider. Our study indicates the possibility of obtaining a clear signal of these new particles with a satisfactory number of events.

\acknowledgements 

One of us (M. D. T.) would like to thank the Instituto de F\'\i sica Te\'orica of the UNESP, for the use of its facilities and the Funda\c c\~ao de  Amparo \`a Pesquisa no Estado de S\~ao Paulo (Processo No. 99/07956-3) for full financial support.


\newpage

\begin{center}
FIGURE CAPTIONS
\end{center}


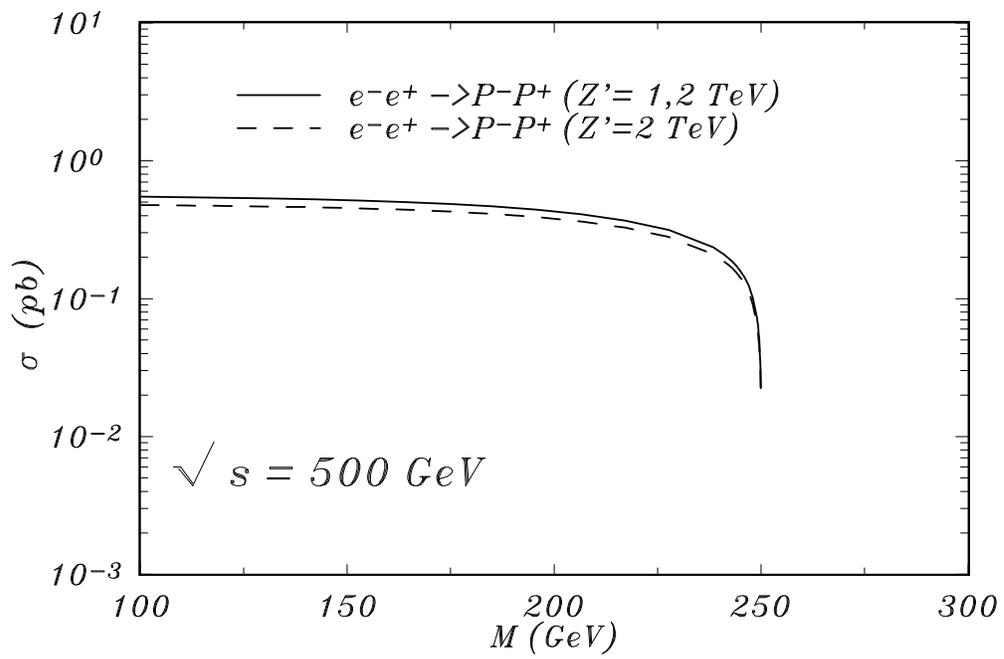
{\bf Figure 1}: Total cross section for the process $e^{-} e^{+} \rightarrow P^-P^+$ as a 
function of $M_{P}$ at $\sqrt{s} = 500$ GeV: (a) $M_{Z'} = 1200$ GeV (solid line), and  (b) 2000 GeV (dashed line).  


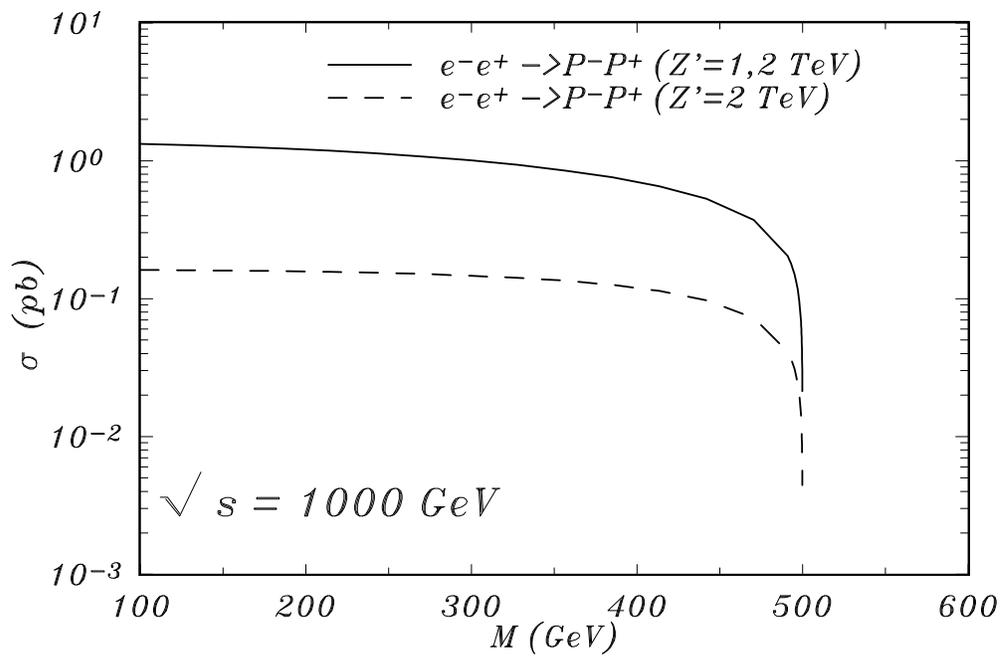
{\bf Figure 2}: Total cross section for the process $e^{-} e^{+} \rightarrow P^-P^+$ as a 
function of $M_{P}$ at $\sqrt{s} = 1000$ GeV: (a) $M_{Z'} = 1200$ GeV (solid line), and  (b) $M_{Z'} = 2000$ GeV (dashed line).  


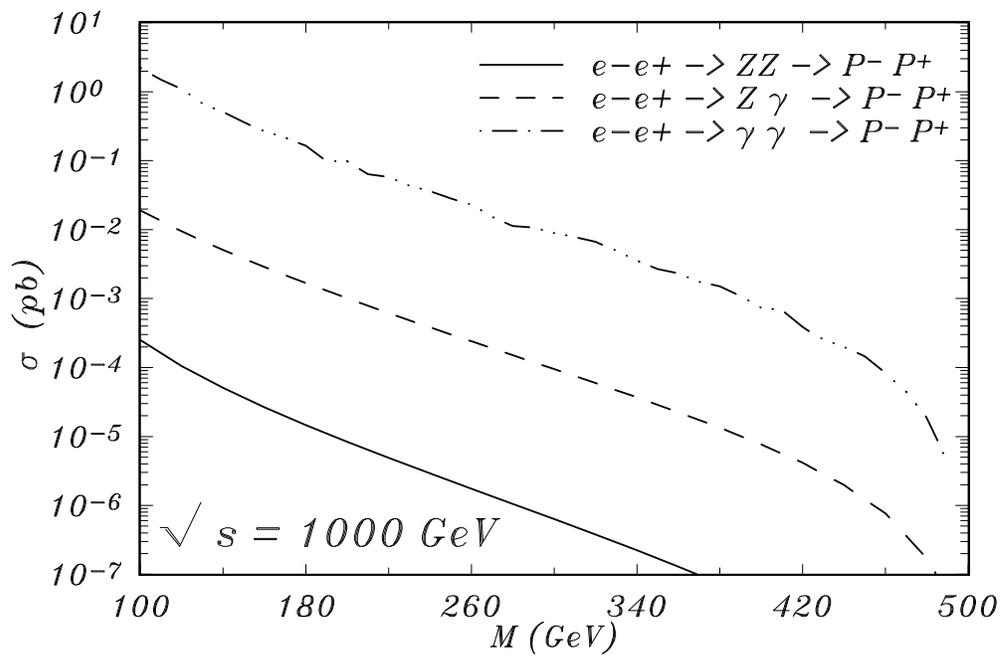
{\bf Figure 3}: Total cross section for the process $e^{-} e^{+} \rightarrow P^{-} P^{+}$ as a function of $M_{P}$ at $\sqrt{s} = 1000$ GeV for different elastic production mechanisms: (a) $\gamma \gamma$ (dot-dot-dashed line); (b) $Z\gamma$ (dashed line); (c) $Z Z$ (solid). 


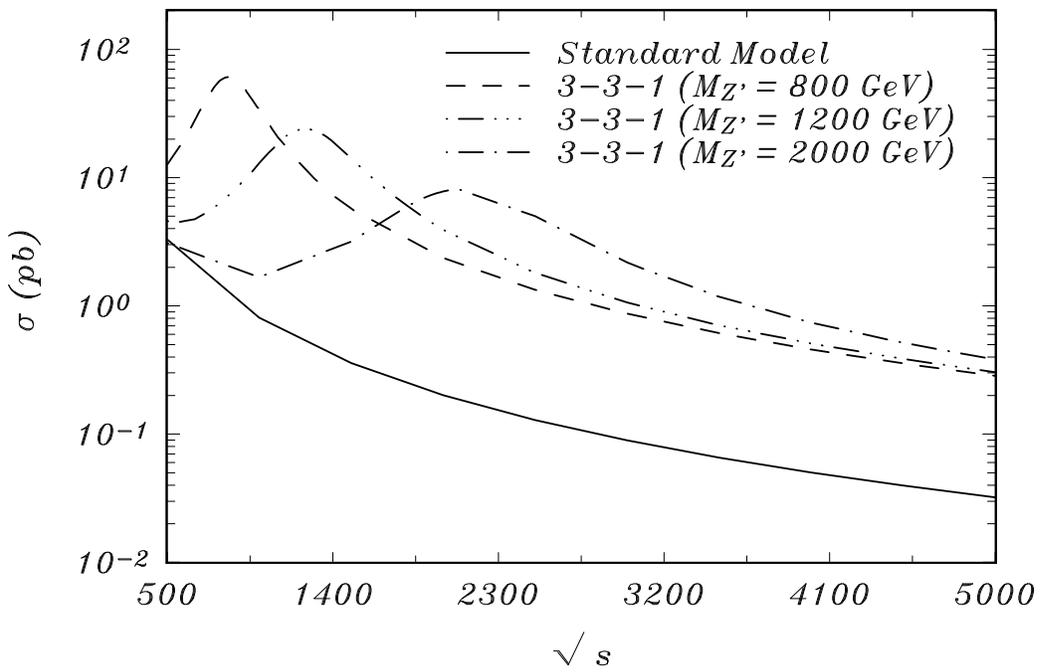
{\bf Figure 4}: Total cross section versus the total c.m. energy $\sqrt{s}$ for the following masses of the gauge boson $Z^{'}$ (a) $M_{Z'} = 800$ GeV (dashed line), (b) $M_{Z'} = 1200$ GeV (dot-dot-dashed line), (c) $M_{Z'} = 2000$ GeV (dot-dashed line), (d) standard model (solid line). 

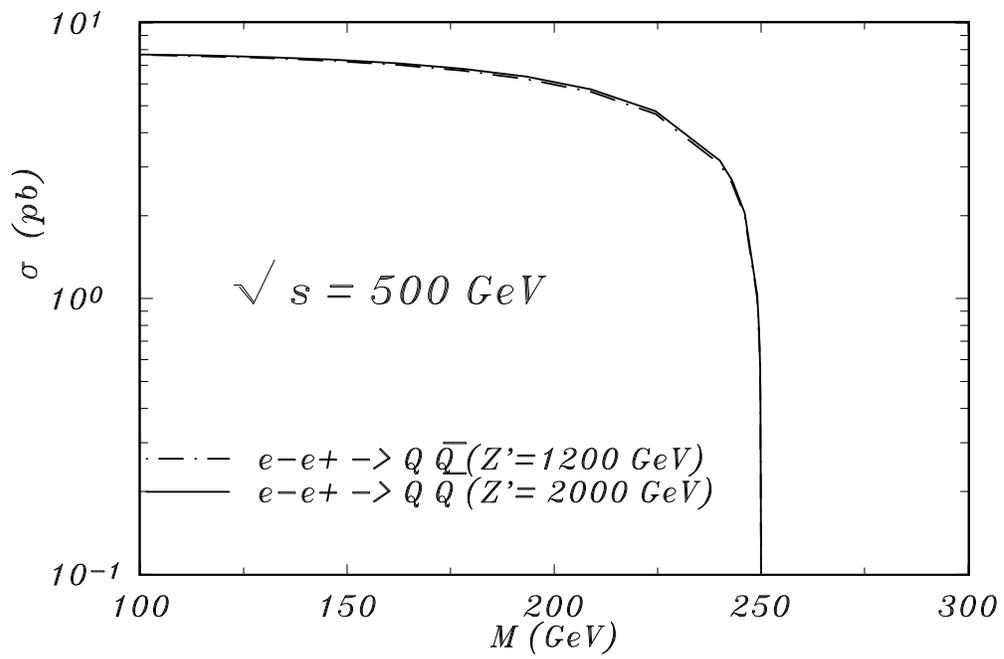
{\bf Figure 5}: Total cross section for the process $e^{-} e^{+} \rightarrow Q \bar{Q}$ as a function of $M_{Q}$ at $\sqrt{s} = 500$ GeV: (a) $M_{Z'} = 1200$ GeV (dot-dashed line), (b)$M_{Z'} = 2000$ GeV (solid line). 


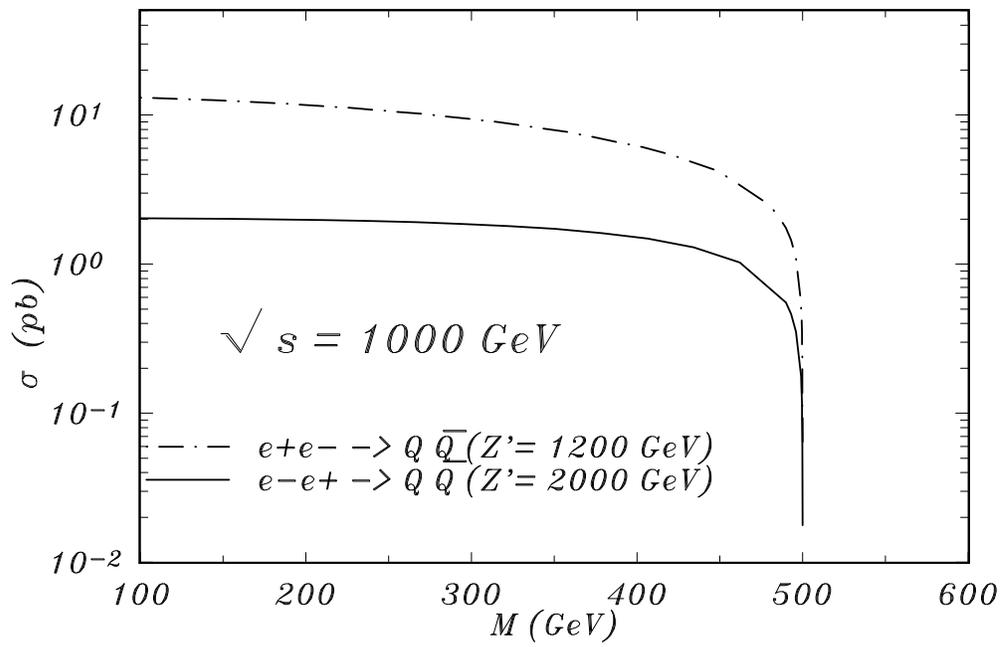
{\bf Figure 6}: Total cross section for the process $e^{-} e^{+} \rightarrow Q \bar{Q}$ as a function of $M_{Q}$ at $\sqrt{s} = 1000$ GeV: (a) $M_{Z'} = 1200$ GeV (dot-dashed line), (b)$M_{Z'} = 2000$ GeV (solid line). 


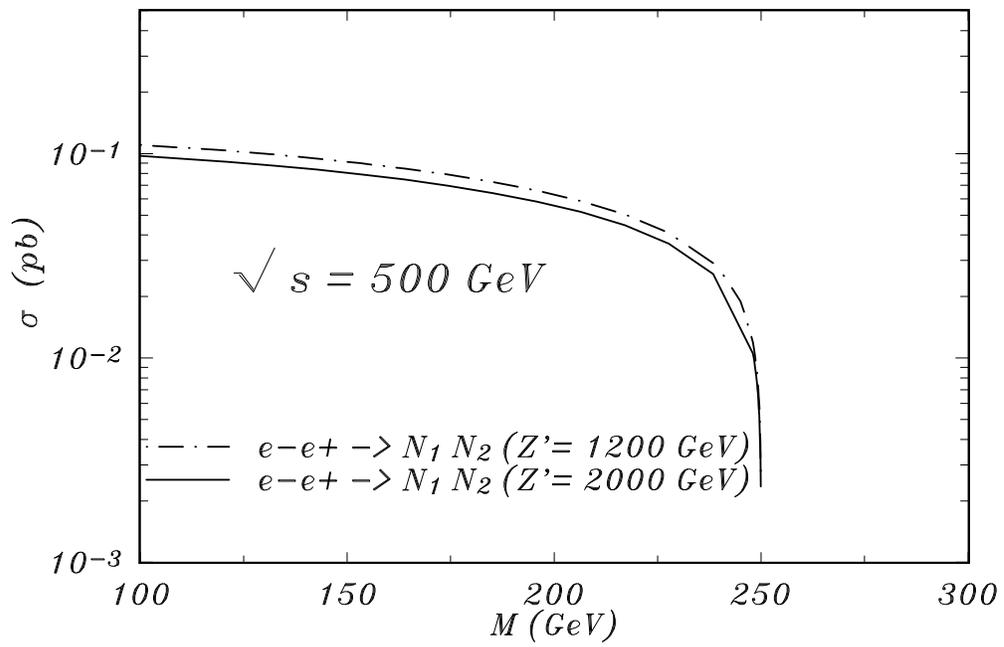
{\bf Figure 7}: Total cross section for the process $e^{-} e^{+} \rightarrow N_{1} N_{2}$ as a function of $M_{N}$ at $\sqrt{s} = 500$ GeV: (a) $M_{Z'} = 1200$ GeV (dot-dashed line), (b)$M_{Z'} = 2000$ GeV (solid line). 


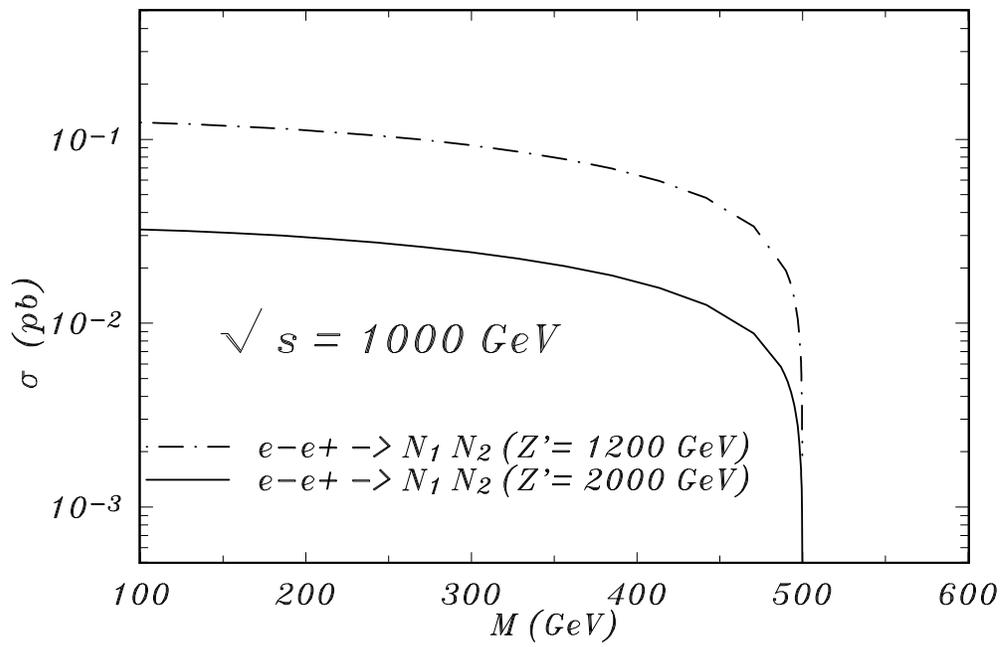
{\bf Figure 8}: Total cross section for the process $e^{-} e^{+} \rightarrow N_{1} N_{2}$ as a function of $M_{N}$ at $\sqrt{s} = 1000$ GeV: (a) $M_{Z'} = 1200$ GeV (dot-dashed line), (b)$M_{Z'} = 2000$ GeV (solid line). 


\end{document}